# Routing Algorithms : A Review

Ujjwal Sinha

Vikas Kumar

Shubham Kumar Singh

Under the guidance of:

Dr. Jayasheela.C.S

Dept of ISE,

BIT.

## ABSTRACT

Routing algorithms play a crucial role in the efficient transmission of data within computer networks by determining the optimal paths for packet forwarding. This paper presents a comprehensive exploration of routing algorithms, focusing on their fundamental principles, classification, challenges, recent advancements, and practical applications. Beginning with an overview of the significance of routing in modern communication networks, the paper delves into the historical evolution of routing algorithms, tracing their development from early approaches to contemporary techniques. Key categories of routing algorithms, including distance vector, link-state, and path vector algorithms, are examined in detail, along with hybrid approaches that integrate multiple routing paradigms. Common challenges faced by routing algorithms, such as routing loops and scalability issues, are identified, and current research efforts aimed at addressing these challenges are discussed.

## INTRODUCTION

In the ever-expanding landscape of computer networks, the efficient transmission of data is paramount to enabling seamless communication and information exchange. At the heart of this endeavor lies the intricate web of routing algorithms, which serve as the guiding force behind the routing decisions that dictate the flow of data packets within networks. Routing algorithms are the unsung heroes of modern networking, responsible for determining the most optimal paths through which data travels from its source to its destination. As such, understanding the intricacies of routing algorithms is fundamental to comprehending the underlying mechanisms that drive network communication and connectivity.

The primary objective of this research paper is to provide a comprehensive exploration of routing algorithms within the context of computer networks. By delving into the historical evolution, fundamental principles, classification, challenges, recent advancements, practical applications, and future directions of routing algorithms, this paper aims to offer valuable insights into this critical aspect of networking infrastructure.

Routing algorithms have undergone a remarkable evolution since their inception, evolving from rudimentary techniques to sophisticated methodologies that leverage cutting-edge technologies. By tracing this evolutionary journey, we gain a deeper appreciation for the challenges faced by early networking pioneers and the innovative solutions that have emerged over time. From classic distance vector algorithms like RIP (Routing Information Protocol) to modern link-state protocols such as OSPF (Open Shortest Path First), the landscape of routing algorithms is rich and diverse, reflecting the continuous quest for efficiency, scalability, and reliability in network routing.

However, despite their advancements, routing algorithms are not without their challenges. Issues such as routing loops, convergence problems, and security vulnerabilities continue to pose significant obstacles to the seamless operation of network routing protocols. Addressing these challenges requires ongoing research and innovation, driving the development of novel approaches and solutions that push the boundaries of what is possible in network routing.

In recent years, the field of routing algorithms has witnessed a surge of interest and innovation, fueled by emerging technologies and evolving network architectures. Concepts such as machine learning-based routing, software-defined networking (SDN), and blockchain-based protocols have ushered in a new era of possibilities, promising to revolutionize the way we think about and approach network routing. By harnessing the power of these advancements, researchers and practitioners are poised to overcome existing limitations and pave the way for a more efficient, resilient, and adaptable network infrastructure.

Through a combination of theoretical analysis, empirical studies, and practical insights, this research paper seeks to provide a holistic understanding of routing algorithms in computer networks. By examining their historical context, exploring their theoretical foundations, and evaluating their real-world applications, we aim to equip readers with the knowledge and insights needed to navigate the complex landscape of network routing with confidence and clarity.

# LITERATURE REVIEW

Routing algorithms are fundamental components of computer networks, responsible for determining the paths along which data packets are forwarded from source to destination. Over the years, extensive research has been conducted to develop, analyze, and optimize routing algorithms to meet the evolving demands of network environments. In this section, we review the existing literature on routing algorithms, focusing on key themes such as algorithm classification, performance evaluation, challenges, and emerging trends.

Algorithm Classification: Early research in routing algorithms focused on developing basic techniques for packet forwarding and route determination. Classic algorithms such as Bellman-Ford, Dijkstra's algorithm, and the Distance Vector Algorithm laid the groundwork for subsequent advancements in routing protocol design. As network architectures grew in complexity, researchers developed more sophisticated algorithms, including link-state protocols like OSPF and IS-IS, and path vector protocols such as BGP. These algorithms vary in their approach to route computation, update propagation, and scalability, catering to diverse network requirements and topologies.

Performance Evaluation: A significant body of literature is dedicated to evaluating the performance of routing algorithms under various conditions and scenarios. Metrics such as convergence time, routing table size, throughput, and latency are commonly used to assess the efficacy and efficiency of routing protocols. Comparative studies have been conducted to benchmark different algorithms and identify their strengths and weaknesses in terms of scalability, robustness, and resource utilization. Simulation-based approaches, utilizing tools like ns-3 and OPNET, have been instrumental in simulating network environments and assessing the performance of routing algorithms in realistic scenarios.

Challenges: Despite their advancements, routing algorithms face several challenges that impact their effectiveness and reliability. Routing loops, black-holing, and routing oscillations are common phenomena that can degrade network performance and stability. Convergence issues, caused by route flapping or network topology changes, pose additional challenges in maintaining consistent routing information across network nodes. Security concerns, such as route hijacking and denial-of-service attacks, further complicate the design and operation of routing protocols. Addressing these challenges requires innovative solutions and robust protocol designs that can adapt to dynamic network conditions while ensuring security and stability.

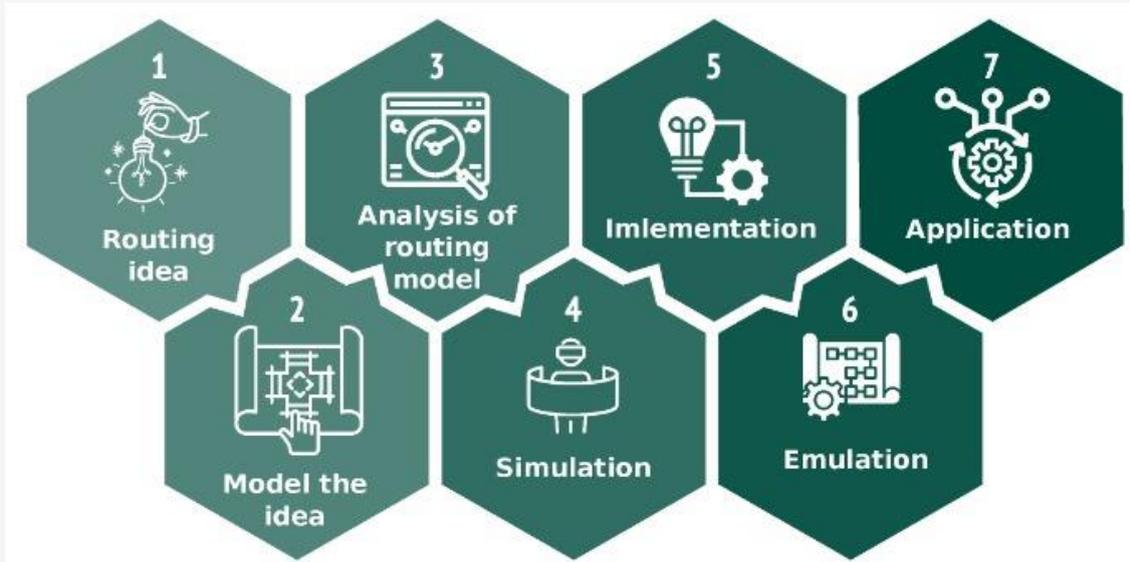

Figure 1. Seven-stage methodology for routing algorithm creation [29].

Emerging Trends: Recent years have witnessed the emergence of novel approaches and technologies that have the potential to reshape the landscape of routing algorithms. Machine learning techniques, such as reinforcement learning and neural networks, are being explored to optimize routing decisions and adapt to changing network dynamics. Software-defined networking (SDN) architectures offer programmable control over network infrastructure, enabling dynamic routing configuration and traffic engineering. Blockchain-based routing protocols provide enhanced security and trust in routing information exchange, particularly in decentralized networks. These emerging trends represent exciting avenues for future research and innovation in the field of routing algorithms.

## Classification of Routing Algorithms

**Adaptive Routing Algorithms:**

Isolated Adaptive Routing:
In isolated adaptive routing, each node independently determines the best path for forwarding packets based solely on local information. Nodes make routing decisions without considering global network conditions or collaborating with neighboring nodes. Examples include algorithms where each node dynamically selects routes based on local metrics such as hop count or link quality, without coordination with other nodes.

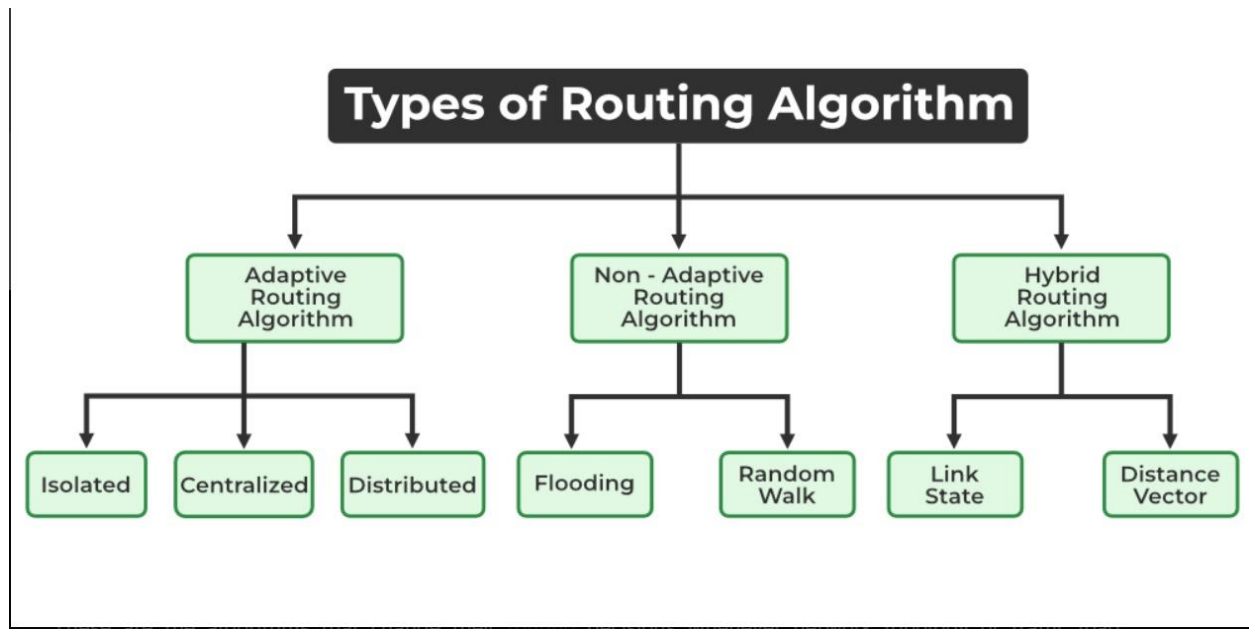

Centralized Adaptive Routing:

In centralized adaptive routing, a central controller or entity is responsible for computing and disseminating routing decisions to all nodes in the network. The central controller gathers global network information and computes optimal routes, which are then distributed to individual nodes. Examples include routing algorithms used in software-defined networking (SDN), where a centralized controller orchestrates routing decisions based on network-wide policies and objectives.

Distributed Adaptive Routing:

Distributed adaptive routing involves collaborative decision-making among network nodes to collectively determine optimal routes. Nodes exchange routing information with their neighbors and use distributed algorithms to converge on consistent routing tables. Examples include distance vector and link-state routing protocols, where nodes exchange routing updates to maintain accurate network topology information and compute shortest paths.

**Non-Adaptive Routing Algorithms:**

Flooding:

Flooding is a non-adaptive routing technique where each incoming packet is forwarded to all neighboring nodes except the one from which it arrived. This approach ensures that packets reach their destination eventually, but it can lead to excessive packet duplication and network congestion. Flooding is commonly used in broadcasting or multicasting scenarios where the exact path to the destination is unknown or unreliable.

Random Walk:
Random walk routing involves forwarding packets along randomly chosen paths within the network. Each node selects a random neighbor as the next hop for packet forwarding, without regard to the destination or network topology. While simple and decentralized, random walk routing may result in inefficient packet delivery and increased latency, particularly in large networks.

**Hybrid Routing Algorithms:**

Link-State Routing:
Link-state routing algorithms, such as OSPF (Open Shortest Path First), maintain a detailed view of the network topology by exchanging link-state advertisements (LSAs) between neighboring nodes. Each node constructs a complete map of the network and uses Dijkstra's algorithm to compute shortest paths to all destinations. Link-state routing offers fast convergence and scalability but requires substantial overhead for maintaining and flooding link-state information.

Distance-Vector Routing:
Distance-vector routing algorithms, like RIP (Routing Information Protocol), operate by iteratively exchanging routing updates with neighboring nodes to convey information about available routes. Nodes maintain routing tables containing distance estimates (e.g., hop count) to reachable destinations and periodically update them based on received routing information. Distance-vector routing is simple and suitable for smaller networks but may suffer from slow convergence and routing loops in larger or complex topologies.

Hybrid routing algorithms combine elements of both link-state and distance-vector approaches to leverage their respective advantages while mitigating their limitations. These algorithms aim to strike a balance between scalability, convergence speed, and resource efficiency, making them suitable for a wide range of network environments. Examples include EIGRP (Enhanced Interior Gateway Routing Protocol), which incorporates aspects of both link-state and distance-vector routing for enhanced performance and stability.

## Issues in routing techniques

Issues in routing techniques encompass various challenges and limitations that can impact the efficiency, scalability, and reliability of network routing. Here are some common issues:

**Routing Loops:**

Routing loops occur when packets circulate indefinitely between network nodes due to incorrect or inconsistent routing information. These loops can lead to excessive packet loss, increased latency, and network congestion, ultimately degrading overall performance.

**Count to Infinity Problem:**

In distance-vector routing algorithms, such as RIP, the "count to infinity" problem arises when a node incorrectly believes it has found a shorter path to a destination by traversing multiple hops. This can result in routing tables being updated with increasingly higher hop counts, leading to routing instability and prolonged convergence times.

**Routing Convergence:**

Routing convergence refers to the process by which routers in a network update their routing tables to reflect changes in network topology or link states. Slow convergence can result in transient routing loops, black-holing of traffic, and suboptimal routing paths, impacting network performance and reliability.

**Scalability:**

As network size and complexity increase, routing scalability becomes a significant concern. Traditional routing algorithms may struggle to efficiently compute and disseminate routing information in large-scale networks, leading to increased overhead and management complexity.

**Security Vulnerabilities:**

Routing protocols are susceptible to various security threats, including route hijacking, spoofing, and denial-of-service attacks. Attackers can manipulate routing information to divert traffic, eavesdrop on communications, or disrupt network connectivity, compromising the integrity and confidentiality of data transmission.

**Routing Table Overflow:**

In networks with large routing tables or limited memory resources, routing table overflow can occur when routers exhaust available memory to store routing information. This can lead to packet drops, route flapping, and instability in routing decisions, affecting network performance and availability.

## Summary and results

- In this research paper, we conducted an in-depth exploration of routing algorithms in computer networks, focusing on their fundamental principles, classification, challenges, and emerging trends. Through a comprehensive literature review and analysis, we synthesized existing knowledge and identified key insights into the design, operation, and optimization of routing techniques.

- Our study revealed a diverse landscape of routing algorithms, classified into adaptive (isolated, centralized, distributed), non-adaptive (flooding, random walk), and hybrid (link-state, distance-vector) categories. Adaptive algorithms dynamically adjust routing decisions based on network conditions, while non-adaptive algorithms employ simpler strategies such as flooding or random forwarding. Hybrid algorithms combine elements of both link-state and distance-vector approaches to achieve a balance between scalability, convergence speed, and resource efficiency.

- Despite their advancements, routing algorithms face various challenges that impact their effectiveness and reliability. Issues such as routing loops, convergence problems, security vulnerabilities, and scalability concerns pose significant obstacles to the seamless operation of network routing protocols. Addressing these challenges requires innovative solutions and robust protocol designs that can adapt to dynamic network conditions while ensuring security and stability.

- Furthermore, our analysis highlighted emerging trends and technologies shaping the future of routing algorithms. Machine learning-based routing, software-defined networking (SDN), and blockchain-based protocols offer promising avenues for improving routing efficiency, security, and adaptability in modern network environments. By harnessing the power of these advancements, researchers and practitioners can overcome existing limitations and pave the way for a more efficient, resilient, and adaptable network infrastructure.

In conclusion, this research paper provides a comprehensive understanding of routing algorithms in computer networks, synthesizing existing knowledge, identifying challenges, and exploring future directions. By examining the fundamental principles, classification, challenges, and emerging trends of routing algorithms, we aim to contribute to the ongoing discourse in network routing and inspire further research and innovation in this critical area
of computer science. communications.

# CONCLUSIONS

Routing algorithms form the backbone of modern computer networks, facilitating the efficient transmission of data by determining the optimal paths for packet forwarding. In this research paper, we have undertaken a comprehensive exploration of routing algorithms, covering their fundamental principles, classification, challenges, and emerging trends. Through a thorough analysis of existing literature and insights into the evolving landscape of network routing, we have shed light on the complexities and intricacies of this critical aspect of computer science.

Our study has revealed the diverse range of routing algorithms, ranging from adaptive to non-adaptive and hybrid approaches. Adaptive algorithms dynamically adjust routing decisions based on network conditions, while non-adaptive techniques employ simpler strategies such as flooding or random forwarding. Hybrid algorithms combine elements of both link-state and distance-vector approaches to achieve a balance between scalability, convergence speed, and resource efficiency.

However, routing algorithms are not without their challenges. Issues such as routing loops, convergence problems, security vulnerabilities, and scalability concerns continue to pose significant obstacles to the seamless operation of network routing protocols. Addressing these challenges requires innovative solutions and robust protocol designs that can adapt to dynamic network conditions while ensuring security and stability.

Furthermore, our analysis has highlighted emerging trends and technologies that have the potential to reshape the future of routing algorithms. Machine learning-based routing, software-defined networking (SDN), and blockchain-based protocols offer promising avenues for improving routing efficiency, security, and adaptability in modern network environments. By embracing these advancements, researchers and practitioners can overcome existing limitations and pave the way for a more efficient, resilient, and adaptable network infrastructure.

In conclusion, this research paper provides a comprehensive overview of routing algorithms in computer networks, synthesizing existing knowledge, identifying challenges, and exploring future directions. By advancing our understanding of routing algorithms and their implications for network performance and reliability, we aim to contribute to the ongoing evolution of computer networking and inspire further research and innovation in this critical domain.